\documentclass[sigconf]{acmart}

\usepackage{varwidth,hyperref,algorithm,algpseudocode,graphics,tabularx,tikz,amsmath,multirow,array,subcaption,todonotes}
\usepackage[simplified]{pgf-umlcd}

\setcopyright{acmcopyright}

\usetikzlibrary{spy,arrows,fit,shapes,calc,graphs,snakes,trees,decorations.pathmorphing}
\usepackage{smartdiagram}	

\begin{document}

\acmDOI{}

\acmISBN{}

\acmArticle{}
\acmPrice{}

\acmConference[]{}{}{}
\acmYear{2018}
\copyrightyear{2018}

\begin{abstract}

Exascale systems will suffer failures hourly. HPC programmers rely mostly on application-level checkpoint and a global rollback to recover. In recent years, techniques reducing the number of rolling back processes have been implemented via message logging. However, the log-based approaches have weaknesses, such as being dependent on complex modifications within an MPI implementation, and the fact that a full restart may be required in the general case. To address the limitations of all log-based mechanisms, we return to checkpoint-only mechanisms, but advocate data-flow-driven recovery (DFR), a fundamentally different approach relying on analysis of the data flow of iterative codes, and the well-known concept of data-flow graphs. We demonstrate the effectiveness of DFR for an MPI stencil code to optimise rollback and reduce the overall energy consumption by 10-12\% on idling nodes during localised rollback. We also provide large-scale estimates for the energy savings of DFR compared to global rollback, which for stencil codes increase as $n^2$ for a process count $n$.
\end{abstract}


\title{Energy-efficient localised rollback after failures via data flow analysis} 

\author{ Kiril Dichev}
\affiliation{
\institution{Queen's University Belfast\\Belfast, United Kingdom}
}
\email{k.dichev@qub.ac.uk}
 
\author{Kirk Cameron}
\affiliation{
\institution{Queen's University Belfast\\Belfast, United Kingdom}
}
\email{k.cameron@cs.vt.edu} 
\author{Dimitrios S. Nikolopoulos}
\affiliation{
\institution{Queen's University Belfast\\Belfast, United Kingdom}
}
 \email{d.nikolopoulos@qub.ac.uk}

\maketitle

\section{Introduction}	

It is widely accepted that compute clusters and supercomputers are transitioning towards systems of millions of compute units to satisfy the requirements of compute-intensive parallel scientific applications.
With this increase in compute components, a proportional decrease in the Mean-Time-Between-Failure (MTBF) across parallel executions will follow \cite{Schroeder2010,Zheng2012}, which would make highly scalable parallel application runs infeasible without integrating resilience.

In this manuscript, we focus on recovery from fail-stop errors, i.e. any failures leading to the unexpected termination of an MPI process and the loss of its data; a node crash is among the possible causes of fail-stop errors.
For such failures, checkpoint/restart (C/R) strategies are commonly used; they introduce time redundancy due to the rollback of execution but require fewer additional resources than resource replication techniques.
The recent advances in fault-tolerant MPI library implementations, such as ULFM~\cite{Bland2013}, have integrated efficient and scalable detection and recovery primitives to allow MPI applications to deal with failures.
C/R with global rollback is the most widely used fault tolerance technique in HPC applications today; in global rollback, all processes roll back to the last globally consistent checkpoint, and continue execution.
Global rollback can be universally applied to all types of application kernels, and its significant advantages in failure-prone executions are widely known. 

It is often unnecessary to roll back all participating processes as in global rollback.
Log-based rollback recovery protocols have explored this property in the past, and have successfully reduced the rollback by finding a more recent consistent cut.
The existing research exploring localised recovery in MPI suggests that its end-to-end benefits lie in more energy-efficient rollback \cite{Guermouche2011}.
However, no quantitative work has ever shown the scale of savings that can be made during localised rollback.

We argue that localised rollback can be achieved without the considerable programming and runtime overhead of log-based techniques within MPI library implementations.
We demonstrate that reducing rollback can be programmed statically for a wide range of applications, and our main driver for such a non-global rollback is the inherent data flow of the application.
Therefore, we call our technique data-flow rollback (DFR).
We also provide concrete experimental evidence that energy savings are indeed achievable with little programming overhead, once DFR is implemented.
Importantly, the energy savings, compared to the common global rollback, can be significant if the application kernels have localised data dependencies.
We show that for codes with neighbourhood data dependencies only, we can save energy in the order of $O(n^2)$ for $n$ processes compared to global rollback, which is a very desirable property in the exascale computing era.

Our work differs from existing MPI-based contributions on localising rollback in a number of key aspects.
DFR logs no messages, saving space at runtime, and makes no assumptions that an application or a runtime will provide any message logging capabilities, instead relying entirely on recompute of work, similar to global rollback.
DFR is coupled to the application, and decoupled from the underlying runtime; as a consequence, DFR can be applied to different parallel runtimes, such as MPI implementations without message logging capabilities, or alternatively to a global address space runtime.
DFR also minimises rollback on an algorithmic level, which makes it robust and applicable to any physical topology. In contrast, partial log-based protocols minimise rollback by explicitly specifying the underlying cluster topology, and optimising for it.
Non optimised log-based protocols, on the other hand, introduce the largest logging overhead.

The main contributions are as follows:
\begin{itemize}
\item We describe a non-global rollback mechanism for applications, based entirely on their data flow graphs
\item We implement DFR for a popular MPI Jacobi code, and couple it with a CPU frequency scaling technique
\item We develop a data-flow based simulator for large-scale runs
\item We measure and model the energy savings for the proposed DFR technique, compared to global rollback
\end{itemize}

The paper is organised as follows.
In Sect. \ref{sec:dfr} we introduce the reader to the concept of data-flow rollback.
We then implement the introduced concepts in an MPI code in Sect. \ref{sec:mpi-jacobi}.
We then present our experimental settings, including a small-scale cluster and a simulator, in Sect. \ref{sec:settings}.
The experimental results are detailed in Sect. \ref{sec:evaluation}.
We further model the energy savings rate in Sect. \ref{sec:model}.
In Sect. \ref{sec:related-work} we summarise the related work and position DFR within the fault tolerance domain.
We conclude with Sect. \ref{sec:conclusion}.

\section{Data-flow-driven rollback}
\label{sec:dfr}
\subsection{Summary and Motivation}

\begin{figure}
\centering
\includegraphics[width=0.4\textwidth]{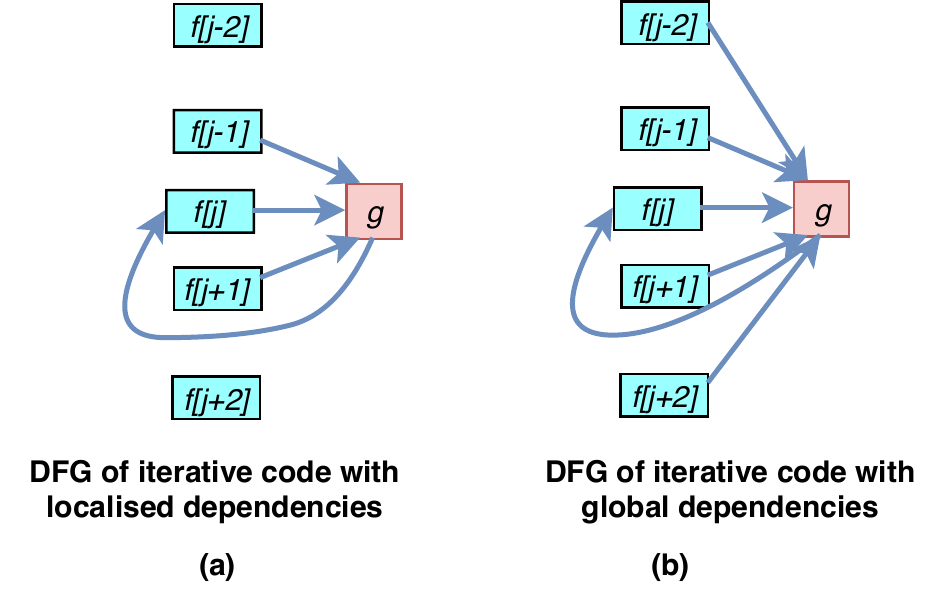}
\caption{Illustration of data flow graphs (DFG) as main motivator for data-flow-driven recovery. The underlying idea is to exploit localised rollback, based on data flow dependencies of the DFG. (a) DFG with localised dependencies and beneficial with DFR. (b) DFG with global dependencies, not beneficial with DFR.}
\label{fig:intro}
\end{figure}

Every iterative algorithm, with timeline ranging from $0 \dots \infty$, can be described as a data flow graph \cite{Parhi2007,Lee1987}, which is a directed graph where the nodes represent units of computation, and the edges represent communication paths.
Consider the illustration given in Fig. \ref{fig:intro}, which summarises how our understanding of iterative applications can educate our decisions on rollback recovery.
Two different data-flow graphs are displayed, which describe how an iteration step is computed. 
$f$ is a partitioned global array, and the function $g$ computes an output out of a number of incoming inputs.
The degree of incoming edges differs for different application kernels.
Fig. \ref{fig:intro}(a) demonstrates that to update partition $f[j]$, we need as input partitions $j-1$, $j$, and $j+1$.
In contrast, for another computational kernel (Fig. \ref{fig:intro}(b)), all the global data may be required for an iteration update.
This observation of data flow is the major driving force for the presented rollback recovery in this work.
In particular, we explore kernels with DFGs that show localised, rather than global, data dependencies.
We use the terms \textit{localised} and \textit{non-global} rollback interchangeably here, referring to any rollback that requires less processes to roll back than in the global rollback case.

In formal terms, each data partition $j$ requires a range of data partitions as input to perform an iteration update $i \rightarrow i +1$:
\begin{equation}
f^{i+1}_j = g(range(f^i_j))
\end{equation}
$g$ is a function performing a computation on its input data, while $range$ returns the range of partitions required to update any given partition $j$. 
The $range$ amounts to the number of incoming arrows into the function block $g$ for the DFG shown in Fig. \ref{fig:intro}.
The data-flow-driven rollback is more beneficial if the indegree of $g$ is small, relative to the number of partitions.

\begin{figure}
\centering
\includegraphics[width=0.4\textwidth]{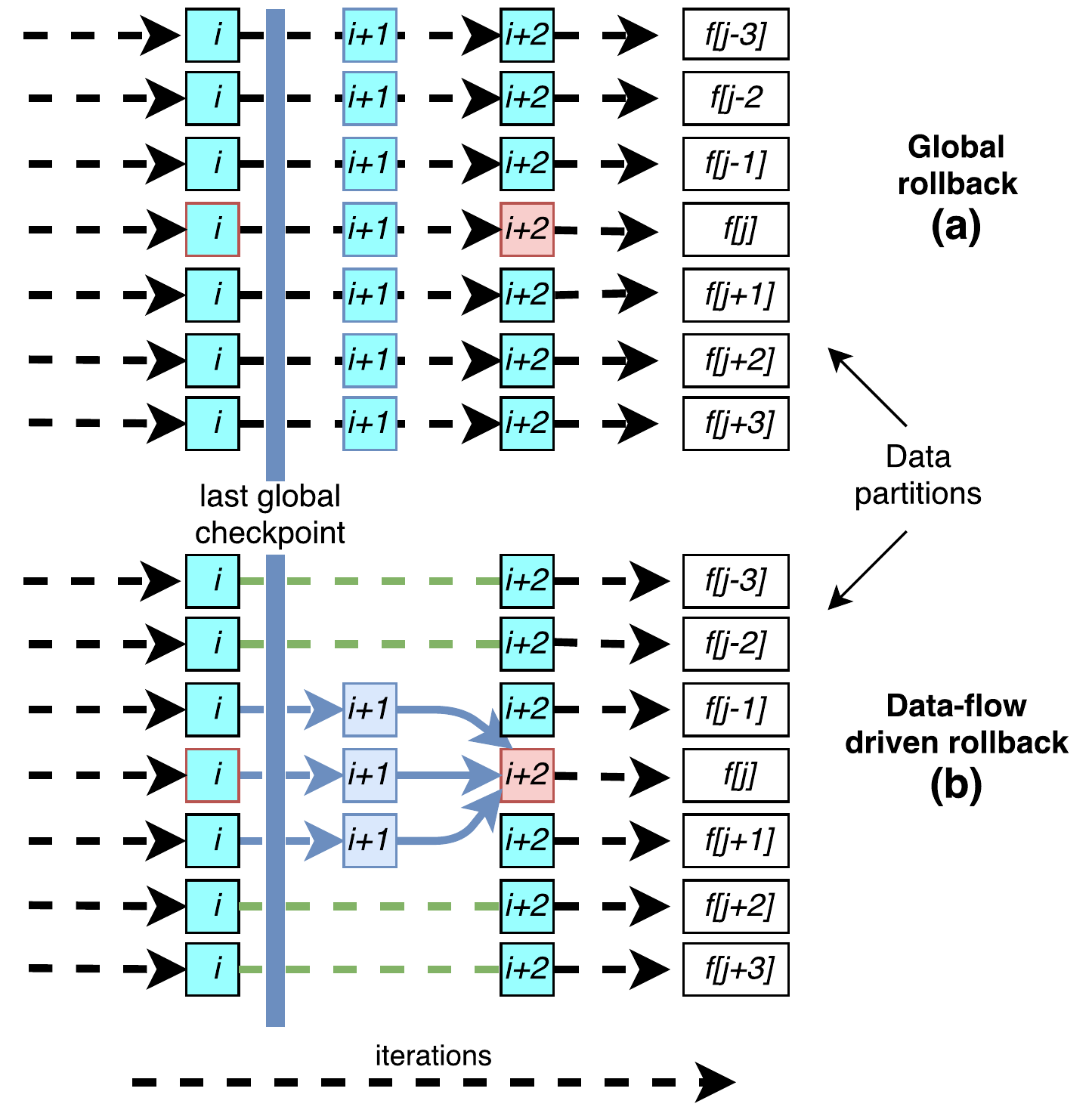}
\caption{Illustration of global and data-flow-driven rollback for codes with neighbourhood dependencies: Marked in red is the process (or data partition) which needs to be recovered. A data-flow-driven rollback releases all processes handling remote data from the need to do rollback.}
\label{fig:ng-rec}
\end{figure}

For example, consider a $range$ operator defined as $range(f_j) = \{f[j-1], f[j], f[j+1]\}$ for the global array $f$; 
this matches the DFG illustrated in Fig. \ref{fig:intro}(a).
An iterative execution for this kernel can experience a process failure, and depending on the rollback mechanism, one of two different execution graphs are possible.
These are illustrated in Fig. \ref{fig:ng-rec}.
In either case, a globally consistent checkpoint is taken after iteration $i$.
Two iterations later, in iteration $i+2$, a process allocated data partition $j$ fails.
The global rollback scenario is illustrated above -- all processes roll back to their local checkpoints of iteration $i$.
This rollback is robust, and oblivious of any underlying data flow dependencies, and computes more than minimally required.
If we used the data flow graph of the kernel, we would recompute partition $j$ for iteration $i+2$ more efficiently.
We can minimally recover by only involving the close neighbour processes, as illustrated in Fig. \ref{fig:ng-rec}(b).
The failed process is replaced (by a spare or respawned process), and a few processes are required to update its data.
The remaining processes are idle during rollback, potentially saving compute resources and energy.
We call this rollback data-flow-driven rollback (DFR), and it is the centrepiece of this contribution.

After a process failure, data is lost, and global data consistency needs to be carefully examined.
For replacement process and surviving processes, this translates into these questions:
\begin{enumerate}
\item How can a replacement process return to the forefront of computation for its data partition, while involving a minimum number of surviving processes?
\item How can all surviving processes retain their data partitions without rolling back, and keep the data consistency?
\end{enumerate}

\subsection{Replacement process}
In this contribution we assume that the failed process is replaced by a replacement process, either from a set of spare nodes, or by respawning a dead process.
The replacement process can advance from the last checkpoint to the beginning of a failed iteration with support from neighbouring processes.
As illustrated, the exact set of rollback processes depends on:
\begin{itemize}
\item $j$, the index of the lost partition
\item $i$, the index of the failed iteration
\end{itemize}
Both of these can be provided at runtime -- $j$ by ULFM, and $i$ by the application.
All of the required processes for the recovery, except for the replacement process, work on ``duplicate data'' for the recovery.
This duplicate data is loaded from the last global checkpoint.
It is discarded upon recovery completion everywhere, except for the replacement process, which keeps this data as its new partition data $f[j]$.
No other process needs to be involved in the recovery.

\subsection{Surviving processes}
All surviving processes keep their partition data during recovery; even the subset of processes involved in rollback do so with temporary copies of their partition. 
Due to the existing data dependencies between surviving processes, their state remains ``almost'' consistent even after recovery -- with the exception of partitions $f[j-1]$ and $f[j+1]$, which may need to be reset to the beginning of iteration $i+2$.

The main advantage of the illustrated data-flow recovery is that only processes responsible for neighbouring partitions of data, in respect to a lost data partition due to failure, need to support the recovery.
This has far reaching consequences for large-scale runs, and in particular for exascale, since the illustrated recovery is independent of the number of data partitions / processes.
The larger the scale of the application run, the larger the benefits of data-flow rollback compared to the wasteful global rollback.

\section{Rollbacks for MPI-based Jacobi method}
\label{sec:mpi-jacobi}
In this section, we detail the implentation of global rollback and our DFR technique for MPI-based Jacobi implementation, which is representative for stencil codes in 2D.

\subsection{Global rollback for MPI Jacobi method}
Our DFR implementation builds on the global rollback recovery code for a Jacobi solver provided for open access by the ULFM developers in hands-on form and documented \cite{site:ulfm-tutorial}.

The Jacobi iteration, defined as follows \cite{Gropp1999}, can be repeated until convergence is reached:
\begin{equation}
u_{i,j}^{k+1} = \frac{1}{4}(u_{i-1,j}^k+u_{i,j-1}^k+u_{i+1,j}^k+u_{i,j+1}^k-h^2*f_{i,j})
\label{eq:jacobi}
\end{equation}

When parallelising the Jacobi iteration via MPI, domain decomposition is required. 
A common decomposition is the 2D Cartesian decomposition, where the 2D grid is partitioned into 2D subdomains for each process.
Each 2D subdomain holds ghost regions, which need to be exchanged at each iteration with neighbouring processes of the 2D Cartesian decomposition.

The original algorithm used by the ULFM developers is outlined in Alg. \ref{algo:rollback-stencil} in black.

\setlength{\intextsep}{0pt}
\begin{algorithm}
\begin{algorithmic}[1]
 \State set error handlers
\State build row and column communicators
\If {recover}
\State \color{blue} \Call{DFR}{$failed\_rank, failed\_iter$}
\If {CPU scaling enabled} \label{line:freq-scal}
\State reduce CPU frequency to minimum
\State barrier
\State reset CPU frequency
\EndIf \label{line:freq-scal-end}
 \color{black}
 \EndIf

\Repeat
\State exchange data with neighbors ($om$) \label{line:exchange}
\If {time for chkpt}
\State save checkpoint ($om$) \label{line:write-cp}
\EndIf
\State compute local updates and residual ($nm \leftarrow om$)
\State allreduce the residual with all processes \label{line:allreduce}
\State swap $om$ and $nm$ \label{line:swap}
\Until {convergence (iterations or residual)}

\Function{Global-Rollback}{}
    \State get data from buddy
    \State goto to local computation
\EndFunction
\color{blue}{
\Function{DFR}{$failed\_rank,failed\_iter$}
\If {$ \Call{partition-distance}{failed\_rank,rank} < (failed\_iter - last\_ckpt\_iter)$} \label{line:dfr-check}
    \State read checkpoint \label{line:read-cp}
    \State join recovery communicator
    \State create duplicate data structures
    \State perform $\|failed\_iter - last\_ckpt\_iter -1 \|$ iterations with recovery communicator
    \If {($rank = failed\_rank$)} \label{line:dupl-use}
    \State save duplicate data into $om$
    \Else
    \State discard duplicate data
    \EndIf \label{line:dupl-use-end}
\EndIf
\EndFunction
}
\end{algorithmic}
\caption{Global rollback for stencils as documented for ULFM \cite{site:ulfm-tutorial}, and an outline (marked in blue) of DFR for the same code.}
\label{algo:rollback-stencil}
\end{algorithm}

The code relies on the capability of the ULFM MPI implementation to detect failures during MPI calls.
The MPI communication can be at ghost exchange (line \ref{line:exchange}) and norm (line \ref{line:allreduce}), or during checkpoint operations (line \ref{line:write-cp} and \ref{line:read-cp}).
If either call at any process detects a failure, it revokes the communicator globally in an error handling function.
All processes ultimately enter the error handler, which rebuilds the communicator.
During global rollback, surviving and replacement processes follow the same application logic: the last global checkpoint is loaded, and a rollback to that iteration is forced for all processes.
All computation following the last checkpoint is discarded across all processes.

\subsection{DFR for MPI Jacobi method}

Our incremental modifications to the algorithm are highlighted in blue.
The source code is available under \cite{res-proto-url}.

Eq. \ref{eq:jacobi} determines data flow dependencies between neighbouring elements of a 2D grid in order to progress between consecutive iteration steps.
The 2D grid represents the domain decomposition between MPI processes, and is concerned with data flow between processes.
Data flow within a process is irrelevant to our work, since only messages between processes may survive an unexpected process crash, and modify a subdomain.

These inter-process data flow dependencies determine the entire DFR method proposed in this section.
Our proposed rollback is different from the global rollback in previous section.
No surviving process rolls back its data.
However, all survivors examine if their data is required by the replacement process for the data-flow driven recovery.
This is expressed for the virtual topology in line \ref{line:dfr-check}, and is based on how data flows from iteration to iteration.
If the data of a surviving process is required, it creates duplicates of the stencil, which live only as long as the recovery continues.
Once the replacement process restores its state, it keeps the computed data, but all supporting processes discard it to resume computation with the pre-failure partition data (lines \ref{line:dupl-use} -- \ref{line:dupl-use-end}).

It is important to highlight that this implements a correct mechanism for Jacobi codes to perform DFR, which is universally applicable across various platforms without further modifications.
Independent of the physical topology, this approach will use a reduced number of recovery processes, which only depends on the frequency of checkpointing.

There are some important questions on the consistency of data when using data-flow-based recovery.
The recompute of the correct data of a replacement process, until it reaches the iteration of failure, is easy to validate.
However, the consistency of data across all surviving processes after the recovery is not guaranteed in general.
Fortunately, the implementation we use supports us in two ways -- the use of an allreduce for the norm computation, and the use of double buffering.

First, note that the global allreduce of line \ref{line:allreduce} used to calculate the global norm eliminates various issues, since this call implicitly guarantees that no two processes will be in different iterations.
Still, even within the same iteration, processes may be in different stages of the iteration, and still hold local partitions building an inconsistent global partition.
For example, some processes may update their ghost regions successfully, while others may lag behind because one of four neighbour processes fails before / during transmission of the ghost region.
In general, this poses significant questions about data consistency.
However, the use of double buffering, and the swap of buffers (line \ref{line:swap}) after the allreduce prevents these inconsistencies. 

To confirm this, we examine if inaccuracies manifest during our approach.
We used as an example the summed squares of the matrix across various iterations.
We observed no difference in the summed squares for test runs with fault-free iterations, global rollback, or the implemented data-flow-driven rollback.
This confirms that the proposed data-flow rollback, as implemented for Jacobi, does not break data consistency.

\subsection{Used frequency scaling technique}

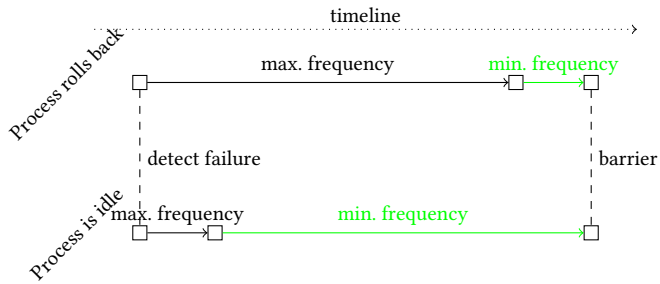
\begin{figure}
\centering
\begin{tikzpicture}[scale=0.4]
\small {
\node[rectangle,draw]  (a) {} ;
\node[rectangle,draw,right of=a, node distance=5cm]  (b) {} ;
\node[rectangle,draw,right of=b, node distance=1cm]  (c) {} ;
\node[rectangle,draw,below of=a,node distance=2cm]  (a2) {} ;
\node[rectangle,draw,right of=a2,node distance=1cm]  (b2) {} ;
\node[rectangle,draw,below of=c,node distance=2cm]  (c2) {} ;
\node[above right of=c] (s2) {};
\node[above left of=a] (s1) {};
\node[left of=a,node distance=0cm,label=left:\rotatebox{45}{\begin{varwidth}{2cm}Process rolls back\end{varwidth}}] (pre-a) {};
\node[left of=a2,node distance=0cm,label=left:\rotatebox{45}{\begin{varwidth}{2cm}Process is idle\end{varwidth}}] (pre-b) {};
\graph[use existing nodes] {
a -> [edge label=max. frequency]b;
b -> [edge label=min. frequency,color=green]c;
a2 -> [edge label=max. frequency]b2;
b2 -> [edge label=min. frequency,color=green]c2;
c --[dashed,edge label=barrier] c2;
a --[dashed,edge label=detect failure] a2;
s1 -> [dotted,edge label=timeline] s2;
};
\draw (a) --  (b);
}
\end{tikzpicture}
\caption{Illustration of CPU frequency scaling used across different nodes for data-flow rollback}
\label{fig:scaling}
\end{figure}

To explore potential energy savings, we use a simple form of CPU frequency scaling, which is shown in lines \ref{line:freq-scal} -- \ref{line:freq-scal-end}, and further illustrated in Fig. \ref{fig:scaling}.
During localised rollback, many processes will be idle, and we can reduce their CPU frequency; as we demonstrate in later section, this results in $\approx 10 \%$ reduction on energy consumption.
We are not aware of other existing energy saving mechanisms for localised rollback.
Indeed, when localising rollback for existing log-based recovery, the runtime makes inference on the latest consistent cut with the help of logs;
it is an open research question if the runtime can also perform frequency scaling in a similar vein.

For the used Haswell processors, the frequency range without voltage scaling (which we do not employ) is between minimum 1.2 GHz and maximum of 3.2 GHz.
Therefore, we cap the maximum frequency of a host CPU of an idling MPI process to 1.2 GHz, and then reset it to the default 3.2 GHz once all processes reach a synchronised post-recovery point.
Note that for the Haswell processors we use the default \textit{intel\_pstate} \cite{intel-pstate-online} module loaded by the Linux kernel (we use CentOS 7 with the default kernel 3.10).
This module has different semantics than the older \textit{cpufreq}, and not all modifications to frequency have the same effect as older modules.
The only modification we employ is capping the maximum frequency of all cores, and then resetting it to its default.
Note that this technique does not increase the overall execution time in our experiments, since it only performs frequency scaling for idle processors.
We have observed that when the frequency is capped recklessly, it can double the execution time, since the frequency is roughly halved from the nominal value of 2.2-2.4 GHz to 1.2 GHz.

\section{Cluster and simulator settings}
\label{sec:settings}
In this section, we provide a summary of the cluster setup for MPI experiments and power measurements, followed by a summary of the simulator we implement for exploring different resilience strategies at scale.

\subsection{Platform Setup}
\label{sec:cluster-platform}
\begin{figure}
\centering
\includegraphics[width=0.35\textwidth]{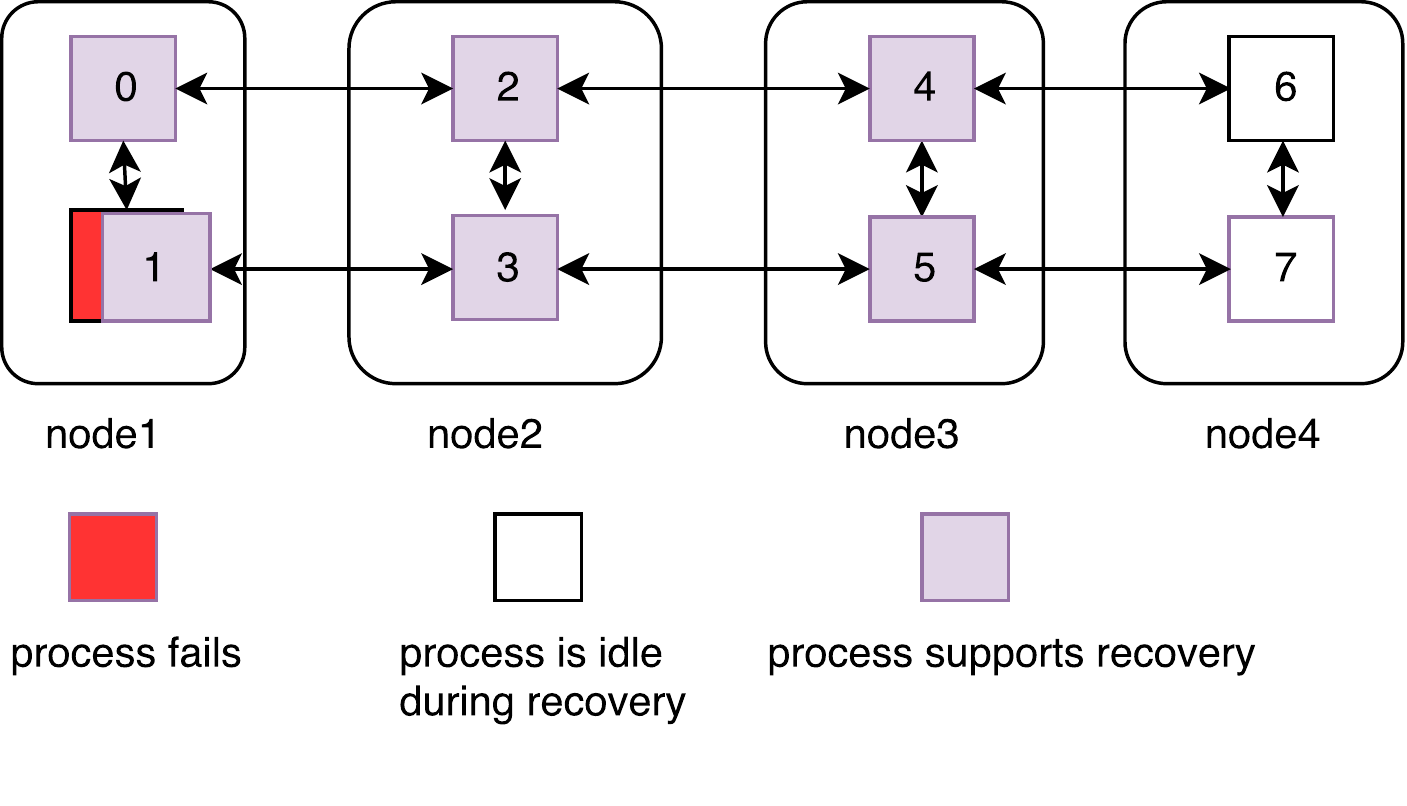}
\caption{Cartesian topology of the 8 MPI process runs of the Jacobi method, mapped onto 4 physical nodes for our experimental setup.}
\label{fig:kos}
\end{figure}

Our platform setup is shown in Fig. \ref{fig:kos}.
We use 4 physical nodes equipped with Intel Haswell processors, with 32 GB RAM each, and connected via 1Gbit Ethernet.
The total power consumption of each physical node is measured by a Sentry Switched PDU.
We sample the power output at each node every 0.5 seconds, using SNMP to read power output at each node.
Since we found no suitable tool to measure and visualise the power consumption at this relatively fine granularity for a PDU, and since we have a specific MPI application setup, we are forced to devise our own measurements.
We sample power consumption on 2 physical nodes of the physical setup shown in Fig. \ref{fig:kos}, with one node participating in non-global recovery (node 3), and one node that remains idle during the non-global recovery (node 4).
We manually filter out all readings of power output less than 110 watts, based on the observation that each node consumes around 100 watts without any workloads running.
We perform each local and global rollback 10 times, and compile the power readings into an aggregated text file.
Each individual data point in the subsequent plots is the mean over 10 iterations.
Due to the manually compiled data and possibility for error, we additionally measure the standard deviation from the mean at any given data point below 4.5 watts; this is not significant given that power varies within a range of 15-20 watts per run.

\subsection{SimGrid-based resilience simulator for 1D stencils}
\label{sec:sim-platform}
There are a number of reasons to explore various localised recovery techniques with a simulator.
On one hand, it is highly desirable to explore simulated large-scale runs with various rollback strategies, in order to increase our confidence on the localised nature of the proposed data-flow rollback, and to compare it with the traditional global rollback.
On the other hand, the exploration of various rollback strategies is important to properly position DFR techniques.

The fact that data flows are at the center of our studies makes it convenient to focus on simulation frameworks which provide data and control flow dependency support.
Our simulator of choice is the open-source framework SimGrid \cite{Casanova2014}, and in particular its SimDag API.
Note that SimGrid has no resilience capabilities;
instead, a developer needs to use its API (in our case: SimDag) to implement resilience capabilities, which is part of our contribution for this work. 
Some of the advantages of SimGrid are:
\begin{itemize}
\item the SimDag interface provides the possibility to express kernel executions as a directed acyclic graph of tasks, and has an API to express data flow and task flow dependencies
\item SimGrid allows to specify the underlying hardware platform used for an experiment, which means we can explore various physical topologies
\item SimGrid has been demonstrated to scale competitively in direct comparison with other simulators (such as LogGOPSim)
\end{itemize}

Our simulator provides various implementations of rollback recovery techniques for one-dimensional codes with neighbourhood dependencies (mostly: stencils).
Since it is execution-driven, each recovery step can be carefully designed and evaluated, which cannot be directly done with trace-driven simulation.
In our implementation, an application run is explicitly defined as a directed acyclic graph (DAG) of tasks.
The DAG can be seen as ``unrolling'' of the DFG of Fig. \ref{fig:intro} into a number of iterations.
Each task corresponds to exactly one iteration of one MPI process; we find that this mapping of MPI code to the more abstract representation as tasks works well.

All tasks are then scheduled in parallel for execution, respecting their data dependencies.
Each rollback recovery scheme is implemented via discarding and rescheduling tasks.
The simulation also necessarily implements a scheduling scheme, which parallelises execution, while at the same time respecting all underlying task dependencies.
The domain decomposition, and virtual topology, are configurable and flexible within the code.
In addition, the physical topology can be specified and used by our simulator, borrowing the features of the underlying SimGrid framework.

Our 1D simulator and the presented rollback strategies are implemented in around 600 lines of C++ code, and freely available \cite{res-proto-url}.
It has been updated to the most recent SimGrid API (3.19) (released in March 2018).

\subsubsection{Simulated cluster}
We use as a reference for simulating stencils the detailed experimental results provided by Tang et al.~\cite{Tang2011} for an efficient single-node shared-memory implementation of stencils.
In the PSA benchmark of reference work, $10^5$ elements over $10^5$ timesteps are computed in 10 seconds, with a sustained speed of 20 GFLOP/s on average for a 12-core Xeon X5650 processor.
Therefore, we model each node to have a peak processing power of 20 GFLOP/s.
We model one task (which represents one iteration update of a node partition) to take 10 seconds, by configuring it as a 200 GFLOPs task (simulator takes FLOPs as input per task).
Also, we model a simple network as a router to which all hosts are directly linked, with no congestion or serial bus in between.
Each host is linked to the router with 1 Gigabit Ethernet (1Gbit bandwidth), with latency of 50 microseconds.

\subsubsection{Simulated data flow for 1D stencils}
Data flow dependencies for 1D stencils are modelled as illustrated in Fig. \ref{fig:intro}(a).
The update of a partition $i$ at iteration $k$ allows data flow into iteration $k+1$ as: $u^k_i \rightarrow u^{k+1}_{i-1}$, $u^k_i \rightarrow u^{k+1}_{i}$, $u^k_i \rightarrow u^{k+1}_{i+1}$ (boundary partitions carry fewer dependencies).
But how do we model the amount of transferred data at each boundary per task?
Since we have a 1D stencil, boundary exchange could be a single double.
However, this would mean we implicitly translate a $10^5 x 10^5$ PSA benchmark into a larger single-timestep problem.
Instead, we model each boundary exchange to amount to exchange of $10^5$ elements.
This follows the notion that each node runs a full 10-second PSA benchmark per one iteration.
This notion is imperfect -- the fine-grained dependencies are thus grouped as 1 large coarse-grained dependency transferring all PSA data at once -- but it still models all coarse-grained inter-node data flow dependencies.
The transferred data per buddy checkpoint is modelled as transferring each process sudomain ($10^5$ elements).

\section{Experimental Evaluation}
\label{sec:evaluation}
In this section, we first experiment with DFR and global rollback for the MPI Jacobi code, and demonstrate that energy savings can be made for DFR, without loss in overall execution time.
Then, we scale up our experimental settings using the resilience simulator, and this provides us with valuable insight about the amount of rollback and the total runtime at larger scale, and for three different rollback strategies.

\subsection{Power measurements for small-scale runs}
\subsubsection{Failure setting for MPI Jacobi code}
We run 2 MPI processes on each physical node, and each run resembles a 2x4 Cartesian topology of 8 MPI processes (Fig. \ref{fig:kos}), with each MPI process computing a grid of $10K^2$ elements.
We retain the efficient buddy checkpointing scheme of the original code, but we modify the buddies to be paired within a physical node, e.g. processes 0 and 1 are checkpointing buddies confined within node 1, etc.
This is important, since non-uniform checkpointing overheads can skew the measurement results.
We limit Jacobi to 10 iterations, and due to the small scale of the experiment, a global checkpoint is taken only at the end of iteration 0.
This block of iterations, enclosed by a checkpoint at the start, is representative of any given block between two global checkpoints, and serves the purpose of evaluating the energy efficiency of DFR.
We forcefully terminate MPI process 1 at the start of the fourth iteration, having completed 3 full iterations.
This is the trigger for rollback recovery (ULFM detects the failure of a process).
We experiment with both global and local rollback.
The global rollback is as implemented in the sample solution provided by the ULFM developers.
All processes roll back to the checkpoint of iteration 0 upon failure, performing 3 pre-failure, and 9 post-failure full iterations.
For the non-global rollback strategy, we use our DFR implementation, where only the closest neighbours participate in the recovery of the failed process.
The distance from the replacement process is 2 (see line \ref{line:dfr-check} in pseudocode), since to reach iteration 3, the replacement process needs a full boundary exchange and local update for iteration 1 and iteration 2.
These can only be provided with the participation of two neighbours in each of 4 directions in 2D.

\subsubsection{Results}
\begin{figure}
\includegraphics[width=0.45\textwidth]{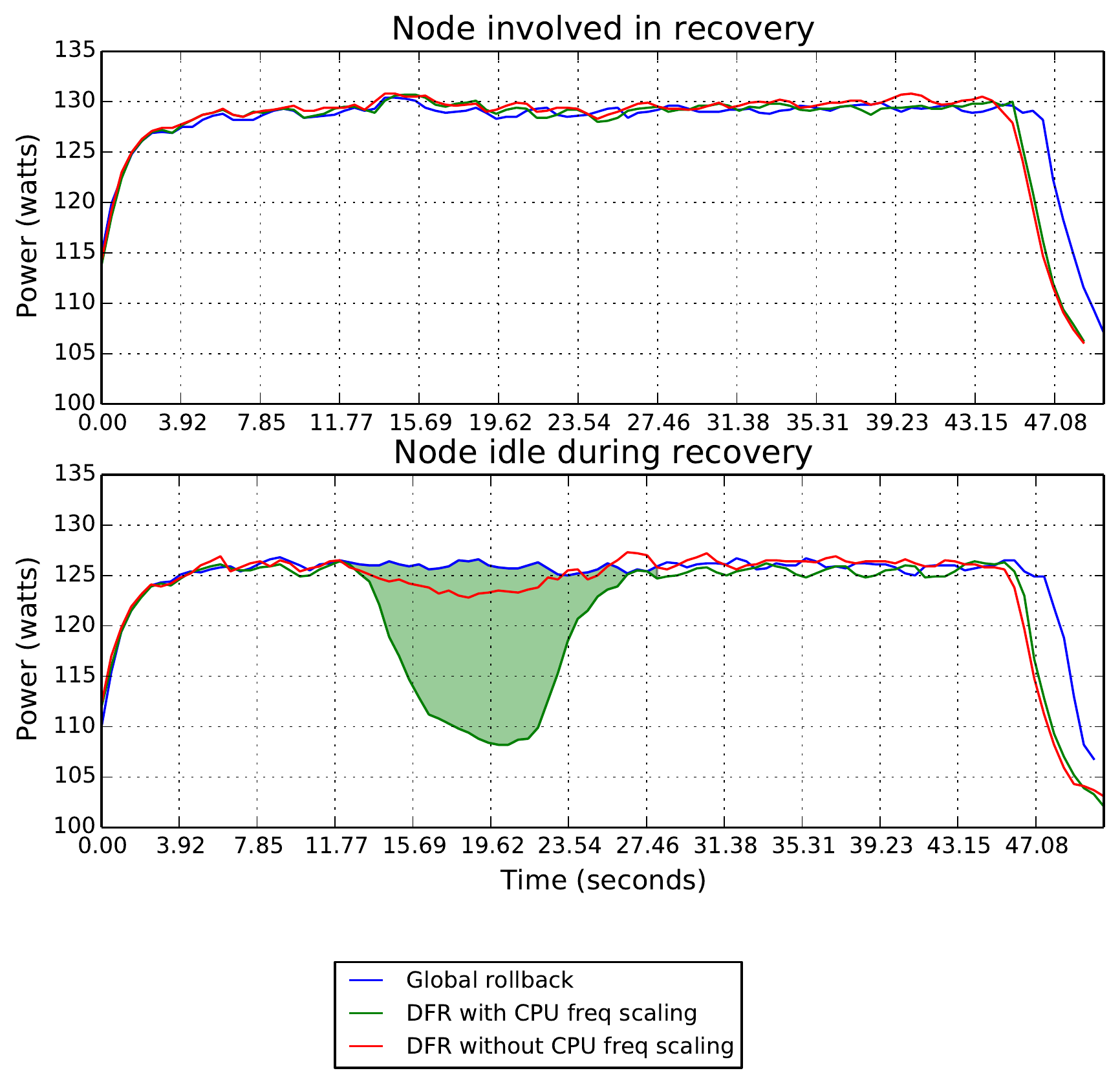}
\caption{Power measurements during global rollback, DFR without CPU scaling, and DFR with CPU scaling.}
\label{fig:power-measurements}
\end{figure}

The results of our experiments are visualised in Fig. \ref{fig:power-measurements}.
The top plot shows the power measurements taken at node 3 of our experimental setup (Fig. \ref{fig:kos}), which participates in the recovery, while the bottom plot shows node 4, which remains idle during the recovery, since its processes 6 and 7 are not needed in the recovery of failed process 1: to recover 2 full iterations, only the closest 2 neighbours are required in the virtual 2D topology.
Each iteration takes $\approx 3.8$ seconds to compute.
We outline the main phases in the execution, in seconds, in case a failure occurs:
\begin{itemize}
\item 0s -- 12s: initialisation, first 3 iterations
\item 12s -- 24s: failure detection, global or local rollback (2 full iterations only on some nodes for DFR, all nodes for global rollback)
\item 24s -- 51s: remaining 7 iterations, termination
\end{itemize}


The global rollback strategy (in blue) involves all processes in equal measures, and shows no drop in energy consumption.
The DFR strategy without frequency scaling (in red), shows a marginal (1-2 J/s) but measurable drop in power consumption during the recovery.
Finally, the DFR strategy with frequency scaling (in green) shows a significant drop in power consumption, from an average of 125 J/s during stencil computation, down to 110 J/s on all idle nodes with reduced frequency scaling.
The curve reflects the significant impact of the introduced frequency scaling technique for idle processors.
Again, we remark that each data point is averaged over 10 iterations of the manually compiled data extracts.

Overall, our evaluation shows that even for this short interval, at least 10-15 J/s, or 10-12 \% of the total energy consumption, can be saved for each idle node per DFR phase.

\subsection{Simulated large-scale runs}
In the simulation experiments, we compare 3 strategies:
\begin{itemize}
\item global rollback for stencils
\item our DFR for stencils
\item our implementation of an efficient stencil-specific log-based recovery, based on work for the S3D application \cite{Gamell2017}\footnote{Upon enquiry, we were unable to obtain the sources of their work}.
\end{itemize}
The presented techniques cover well localised and global rollback (see Fig. \ref{fig:related-work}).
We have implemented support only for one-dimensional stencils for all three rollback techniques in this work.

We only focus on weak scaling experiments --we scale up the node count (assuming 1 MPI process per node), with constant load per MPI process.
We measure various metrics of interest for the same set of experiments.
We do not allow multiple failures at the same time, or during recovery.
We also model detection and replacement process startup as zero-overhead operations, since this study focuses on the amount of rollback.
A total of 1000 iterations are run, and the simulated runtime is $10^4$ seconds, or $\approx$ 2.7 hours.
Each checkpoint transfers its entire stencil array to another host's main memory (buddy checkpointing). 
We use a checkpoint interval of 6 iterations for each strategy; as long as the checkpoint interval is fixed, the relation between techniques is representative.
For each given random seed, different failure times manifest, and the execution differs.
Therefore, we average for each single datapoint over 10 runs with different seeds, using the same seeds for better comparison across rollback strategies.
We set the node MTBF to a rather pessimistic value of 100 hours, with the sole purpose of demonstrating the order of magnitude that different rollback strategies differ in, and since we are unable to scale the experiments to many thousand nodes for that many tasks.

\subsubsection{Recomputed work}
\begin{figure}
\begin{subfigure}{1\columnwidth}
\centering
\includegraphics[width=1\columnwidth]{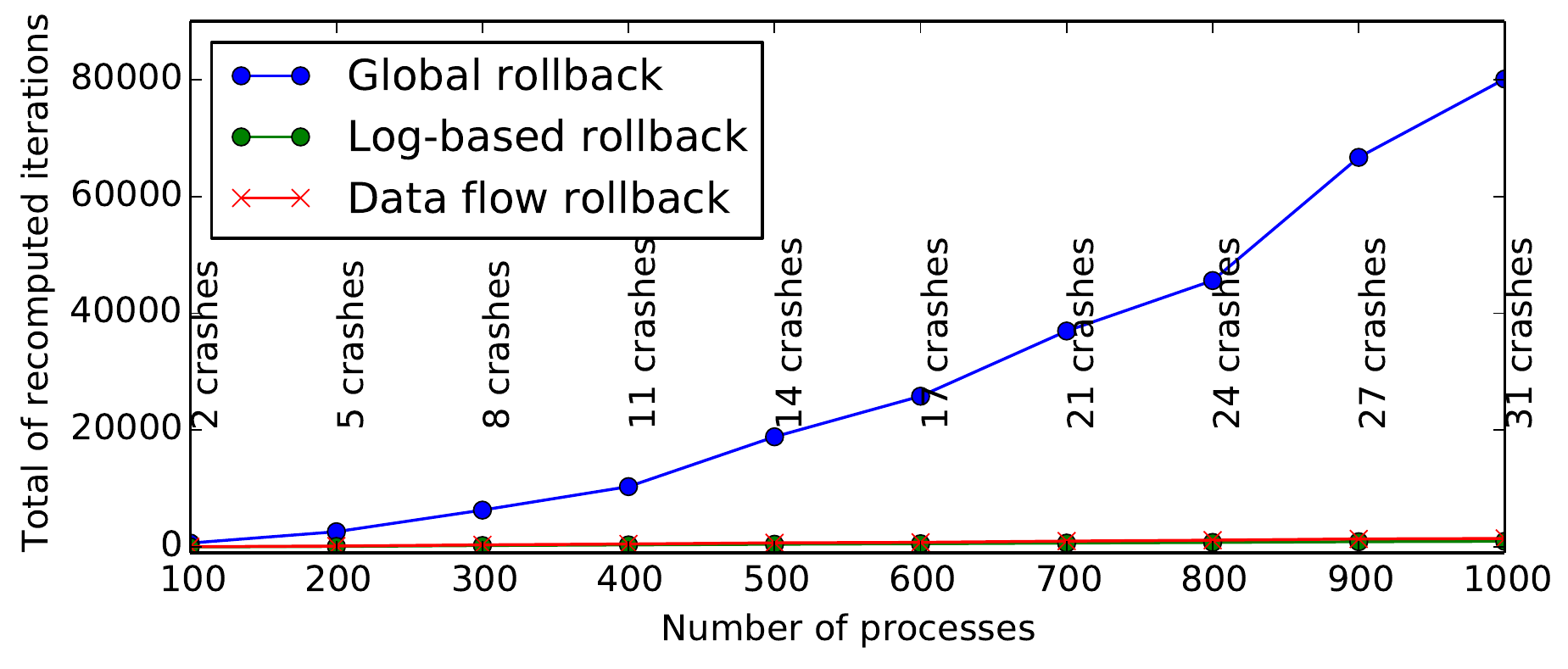}
\caption{Recomputed iterations for global rollback, log-based rollback, and DFR.}
\label{fig:recomputed-tasks-2}
\end{subfigure}

\begin{subfigure}{1\columnwidth}
\centering
\includegraphics[width=1\columnwidth]{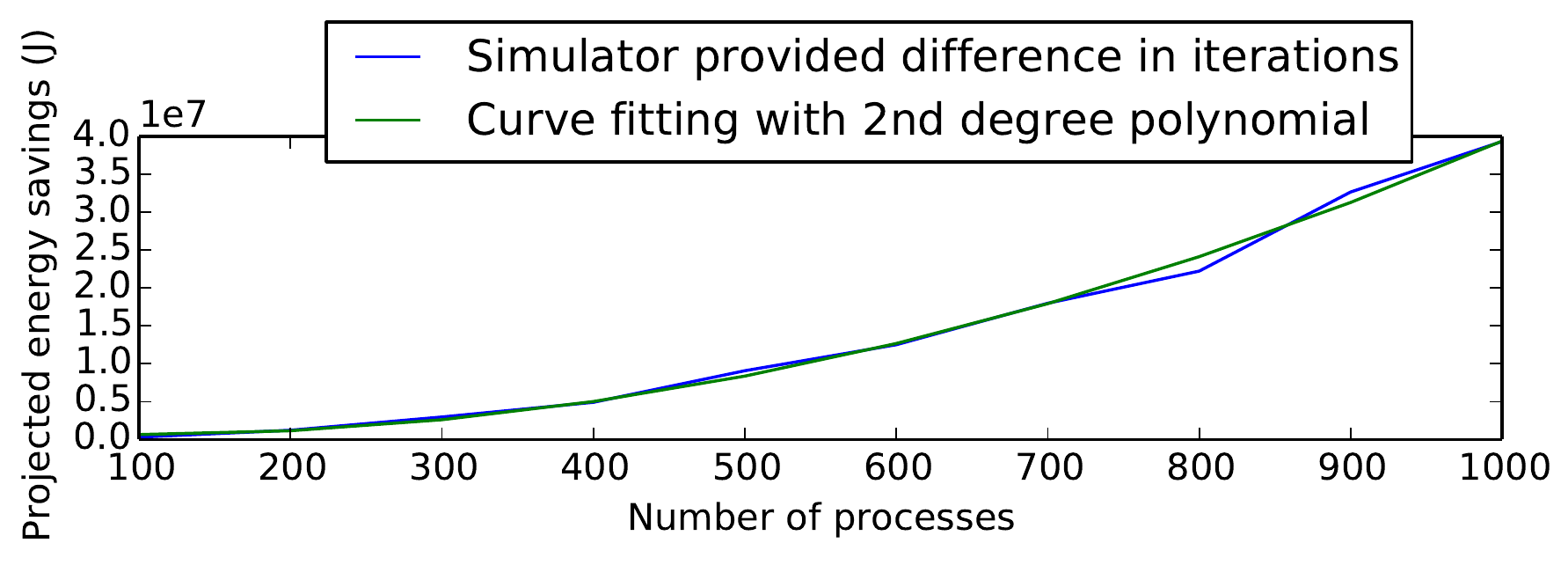}
\caption{Projection of total energy savings (in joules) for DFR compared to global rollback.}
\label{fig:projections}
\end{subfigure}

\begin{subfigure}{1\columnwidth}
\centering
\includegraphics[width=0.8\columnwidth]{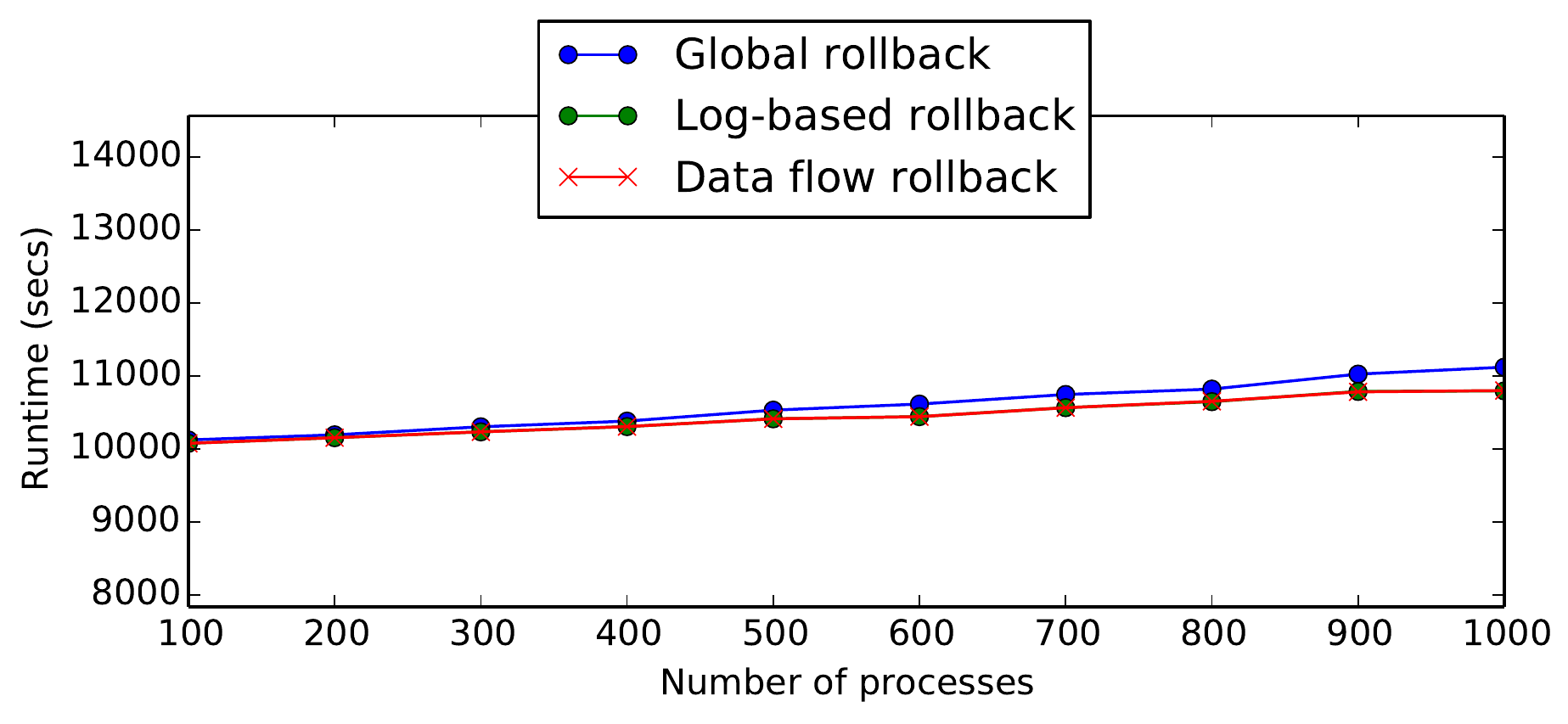}
\caption{Total runtime for global rollback, log-based rollback, and DFR.}
\label{fig:durations}
\end{subfigure}

\caption{1D stencil simulation of $\approx 2.7$ hours, scaled up to 1000 hosts (weak scaling), with MTBF per node of 100 hours.}
\end{figure}
We show the amount of recomputed iterations in Fig. \ref{fig:recomputed-tasks-2}.
The system MTBF linearly increases for larger runs, ranging from 2 crashes for 100 nodes, to 30 crashes for 1000 nodes.
The global rollback strategy recomputes a larger number of tasks than any local recovery technique, rolling back significantly more iterations with increasing MPI process count.
On the other hand, DFR and log-based rollback are both localised in the number of recomputed iterations.
The most efficient of the tested rollback strategies is the highly optimised and stencil-specific log-based rollback, which only needs to replay the lost work on a replacement node; the ghost exchange can be replayed from message logs, and requires no recompute on surviving nodes.
On the other hand, the proposed DFR, which follows the ideas as outlined in previous sections, needs to reschedule work both on the replacement node, and a limited number of neighbouring nodes.
Both the log-based and the data-flow-driven solution, as expected, behave as non-global rollback: the recomputed iterations depend on the checkpointing interval and failure rate, but not on the MPI process count of the parallel runs.
This is an important local recovery property, which we use later (see Eq. \ref{formula:busy-guys} and Eq. \ref{formula:idle-guys}) to model the extent of recompute as non-global.

\subsubsection{Projection of energy savings}
To further quantify the gains of DFR over global rollback in this section, we display in Fig. \ref{fig:projections} the energy savings per run in joules for DFR.
We do this by multiplying the difference in total iterations by a projection of the energy consumed per iteration.
The energy per iteration is here approximated as 500 joules, and adopted from Jacobi experiments, where an iteration of 4 seconds at 125 J/s results in 500 joules (Fig. \ref{fig:power-measurements}).
Then, the savings in energy consumption are in the order of $O(n^2)$, as evidenced by a curve fitting with a second degree polynomial; this is further formalised in later section.
For large-scale runs with 1000 nodes (1 process per host), energy savings of $4*10^7$ joules are projected; note that these significant savings result from the 31 crashes during a 2.7 hour 1000-node run.

\subsubsection{Total runtime}
We also can measure the overall execution time for either of the employed rollback strategies, and show results in Fig. \ref{fig:durations}.
We find any localised rollback, either DFR or log-based, does not show any consistent reduction in overall execution time, compared to global rollback.
This finding is entirely plausible, since not rolling back (for localised rollback) does not imply that execution can freely progress.
The neighbourhood dependencies ultimately propagate to all processes and slow down execution even for localised rollback.
This result served as important hint for us to pursue improvements in overall energy efficiency, rather than overall execution time.
We note that recent research \cite{Gamell2017} shows that overall runtime can be reduced with localised log-based rollback; the authors simulate higher failure rates, where the so called ``failure masking'' is observed.
Our findings of Fig. \ref{fig:durations} do not allow us to claim significant savings in overall execution time for localised rollback approaches.

\section{Model of DFR energy savings for large-scale runs}
\label{sec:model}
We previously demonstrated that using non-global rollback, particularly when combined with simple CPU frequency scaling techniques, yields power savings, compared to global rollback methods; in particular, all processes which are distant from the failed process, (in the virtual topology) can remain idle and scale down the CPU frequency.
Now, we provide an estimate of the rate of overall energy savings across the system for an MPI-parallel run of an application employing DFR, compared to the same run using the global rollback method.
The energy savings are visible through the green timelines illustrated in Fig. \ref{fig:ng-rec}.
There are three parameters to consider for these estimates:
\begin{itemize}
\item The system failure rate $\frac{n}{\mu}$ (unit: $\frac{1}{s}$)
\item The number of idle processes $P_{idle}$ during DFR. This number depends on the DFG of a kernel (no unit)
\item The savings in energy per process and rollback $C_e$; this constant cannot be generalised, and depends on kernel, platform, energy-saving technique etc. (unit: J)
\end{itemize}


We estimate the \textit{rate of energy savings} across the system as the product of all these factors (unit: J/s):
\begin{equation}
E =  \frac{n}{\mu} *  P_{idle} * C_e
\label{formula:savings-rate}
\end{equation}

\subsection{Idle Processes ($P_{idle}$)}
\label{sec:idle-proc}
\begin{figure}
\centering
\begin{tikzpicture}[scale=.4,every node/.style={minimum size=1cm},on grid]
		
    \begin{scope}[scale=0.8,
            ,every node/.append style={
            },
            ]
        \fill[white,fill opacity=0.9] (0,0) rectangle (1,5);
        \draw[step=10mm, black] (0,0) grid (1,5); 
        \draw[step=1mm, red!50,thin] (0,1) grid (1,4);  
        \draw[black,very thick] (0,0) rectangle (1,5);
        
       \begin{scope}[
    	xshift=60,every node/.append style={
    	    },
    	             ]
        \fill[white,fill opacity=.5] (0,0) rectangle (1,5);
        \draw[black,very thick] (0,0) rectangle (1,5);
        \draw[step=10mm, black] (0,0) grid (1,5);
        \draw[step=1mm, red!50,thin] (0,2) grid (1,3);  
    \end{scope}
    
           \begin{scope}[
    	xshift=-60,every node/.append style={
    	    },
    	             ]
        \fill[white,fill opacity=.5] (0,0) rectangle (1,5);
        \draw[black,very thick] (0,0) rectangle (1,5);
        \draw[step=10mm, black] (0,0) grid (1,5);
        \draw[step=1mm, red!50,thin] (0,0) grid (1,5);  
    \end{scope}
    \end{scope}
\begin{scope}[xshift=300,yshift=50,scale=0.75]
    \begin{scope}[
            xshift=-163,every node/.append style={
            xslant=0.5,yslant=-1},xslant=0.5,yslant=-1
            ]
        \fill[white,fill opacity=0.9] (0,0) rectangle (4,4);
        \draw[step=8mm, black] (0,0) grid (4,4); 
        \draw[black,very thick] (0,0) rectangle (4,4);
        
         \draw[step=0.8mm, red!50,thin] (1.6,0) grid (2.4,4.0);  
         \draw[step=0.8mm, red!50,thin] (0,1.6) grid (4,2.4);  
	\draw[step=0.8mm, red!50,thin] (0.8,0.8) grid (3.2,3.2);  
    \end{scope}
    \begin{scope}[
            xshift=-83,every node/.append style={
            xslant=0.5,yslant=-1},xslant=0.5,yslant=-1
            ]
        \fill[white,fill opacity=0.9] (0,0) rectangle (4,4);
        \draw[step=8mm, black] (0,0) grid (4,4); 
        \draw[step=0.8mm, red!50,thin] (1.6,0.8) grid (2.4,3.2);  
        \draw[step=0.8mm, red!50,thin] (0.8,1.6) grid (3.2,2.4);  
        \draw[black,very thick] (0,0) rectangle (4,4);
    \end{scope}
    	
    \begin{scope}[
    	every node/.append style={
    	    xslant=0.5,yslant=-1},xslant=0.5,yslant=-1
    	             ]
        \fill[white,fill opacity=.9] (0,0) rectangle (4,4);
        \draw[black,very thick] (0,0) rectangle (4,4);
        \draw[step=8mm, black] (0,0) grid (4,4);
        \draw[step=0.8mm, red!50,thin] (1.6,1.6) grid (2.4,2.4);  
    \end{scope}
\end{scope}

\path (-3,-1.5) node(x) {} 
      (4,-1.5) node(y) {};
\draw[<-,thick] (x) -- node[label=below:{Rolling back for 1D}] {} (y);
\path (5,-1.5) node(x2) {} 
      (14,-1.5) node(y2) {};
\draw[<-,thick] (x2) -- node[label=below:{Rolling back for 2D}] {} (y2);

\end{tikzpicture}
\caption{Required partitions for a minimal data-flow rollback for 1D 3-point stencils, and for 2D 5-point stencils.}
\label{fig:1d-dep}
\end{figure}
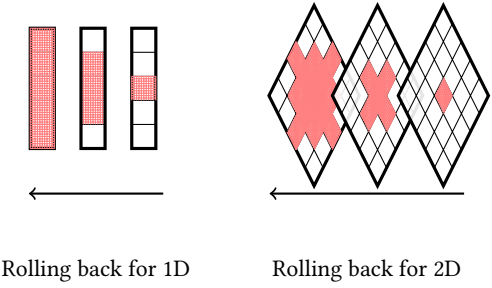

Consider DFR recovery for 1D and 2D stencils, which is illustrated in Fig. \ref{fig:1d-dep} for a minimal and efficient scheme.
An increase in involved processes occurs with each lost iteration; all iterations between the last global checkpoint and the iteration of failure are lost.
If a failure happens in iteration $i$, we extend to the left and right-hand neighbour of the previous iteration $(i-1)$, until we reach the iteration of the last checkpoint.
(Note that the DFR solution for Jacobi shown in Alg. \ref{algo:rollback-stencil} is less efficient than shown for the 2D stencil case of Fig. \ref{fig:1d-dep} for simplicity)

When observing an execution of an arbitrary number of iterations, we only need to study the confined range of iterations between any two global checkpoints; this range is representative of the entire execution, since at any given time the rollback is confined by the last global checkpoint.
Without loss of generality, we only consider that a failure in iteration $i$ happens within the enclosing global checkpoint iterations:  $0 \le i < C_{it}$, where $C_{it}$ is the interval of checkpointing in iterations.
$C{it}$ is the main factor of the remaining estimates of this section.
At any given iteration, only the distance in iterations to the last global checkpoint matters.
Let a failure happen $i$ iterations after the last global checkpoint.
Then the total number of neighbouring processes which need to support the replacement process during non-global recovery is:
\begin{equation}
\begin{split}
P^{1D}_{neigh}(i) = 2*(max(i-1,0)) = \theta(i) \\
P^{2D}_{neigh}(i) = {2*(max(i-1,0))}^2  = \theta(i^2)
\end{split}
\end{equation}

We also estimate the average number of neighbouring processes involved in the recovery, given a checkpoint interval (in iterations) of $C_{it}$.
We consider a uniform probability distribution, i.e. a failure may happen at any given iteration $i$.
We can average the number of active neighbouring processes for a block of iterations enclosed by global checkpoints as:
\begin{equation}
\begin{split}
P^{1D}_{active} = \frac{1}{C_{it}} \sum_{j=1}^{C_{it}} P^{1D}_{neigh}(j) =   \theta(C_{it}) \\
P^{2D}_{active} = \frac{1}{C_{it}} \sum_{j=1}^{C_{it}} P^{2D}_{neigh}(j) =  \theta(C_{it}^2)
\end{split}
\label{formula:busy-guys}
\end{equation}


We now need to inverse this formula to find the number of processes that remain idle during an execution which experiences single failures.
This can then be estimated for $n$ MPI processes as:
\begin{equation}
\begin{split}
P^{1D}_{idle} = n - P^{1D}_{active} \in O(n) \\
P^{2D}_{idle} = n - P^{2D}_{active} \in O(n)
\end{split}
\label{formula:idle-guys}
\end{equation}

Via Eq. \ref{formula:busy-guys}, for 2D stencils, an order of magnitude more processes need to rollback.
Still, as long as we choose a reasonable checkpoint interval, and a large process count $n$ is used, a proportionate number of processes remains idle during localised rollback.
This property is the essence of any localised rollback.

We also note that algorithms with global dependencies may have no idle processors for DFR, which naturally translates to $P_{idle} = 0$ and 0 energy savings, compared to global rollback.

\subsection{System failure rate}
\label{sec:fail-rate}
At exascale, some projections estimate a mean time between failures (MTBF) of hours \cite{Cappello2009}.
There already exists experimental work which has already measured MTBFs of hours \cite{Tiwari2014,Gamell2014}.
Other research also suggests that the MTBF of future systems may be in the order of minutes \cite{Dongarra2011}.

We can estimate the system MTBF for $n$ nodes, assuming the MTBF per node is $\mu$ (independent of other nodes), as $\frac{\mu}{n}$ (e.g. \cite{Bosilca2014}).
The system failure rate is therefore the reciprocal value $\frac{n}{\mu}$.

\subsection{Energy savings per DFR phase and process ($C_e$)}

We experimentally verified in Sect. \ref{sec:evaluation} the energy savings for one MPI application per node , which amount to 10-15 J/s per idle process, or 10-12\% of the total energy consumption, for the duration of the rollback.
To estimate the average duration of recovery, we can apply the same logic as in Sect. \ref{sec:idle-proc}.
Again, we are encapsulated within a global checkpoint interval, and we assume a failure happens with uniform probability in any iteration $i$, with $0 \le i < C_{it}$.
This results in an average of $\frac{C_{it}}{2}$ rollback iterations.
To make things more concrete, we again consider the MPI code we ran, where each iteration runs for nearly 4 seconds.
For energy savings of 10 J/s, this results in average energy savings per DFR phase and process of $C_e = 20 * C_{it}$ J.

\subsection{Overall energy savings for used MPI code and platform}
Based on Eq. \ref{formula:savings-rate}, and the subsequent derivations, we conclude that on the example of the proposed DFR technique for the Jacobi method, our rate of energy savings (J/s) compared to implementing global rollback is:
\begin{equation}
E_{Jacobi} \approx \frac{n^2}{\mu} * 20 * C_{it}
\end{equation}

For example, for an application run with $10^4$ MPI processes (1 process / node), MTBF of nodes of 50 years, and checkpoint interval of 10 iterations, our overall energy savings for the entire run are $\approx 13 $ J/s, as opposed to a strategy employing global rollback for the same code and setup, which is a modest energy saving.
If we instead use e.g. $10^5$ MPI processes (1 process / node), the overall energy savings rate compared to global rollback increases to $13*10^2$ J/s.
This transfers into energy savings comparable to running 13 additional machines for the entire application run, assuming a machine has an energy consumption of 100 J/s.

\section{Related Work}
\label{sec:related-work}

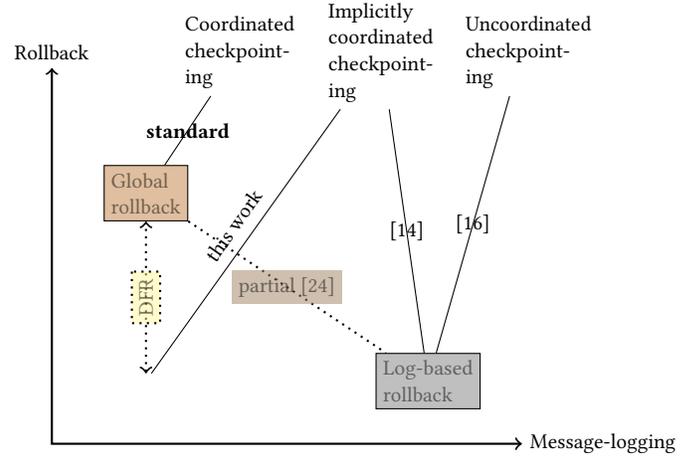
\begin{figure}
\begin{tikzpicture}[scale=2.5]
\small{
    \draw [<->,thick] (0,2) node (yaxis) [above] {Rollback}
        |- (2.5,0) node (xaxis) [right] {Message-logging};
      \node (ucp) at ($(yaxis)+(2.5,0)$) {\begin{varwidth}{1.5cm}Uncoordinated checkpointing\end{varwidth}}; 
      \node (iccp) at ($(yaxis)+(1.75,0)$) {\begin{varwidth}{1.5cm}Implicitly coordinated checkpointing\end{varwidth}}; 
      \node (ccp) at ($(yaxis)+(1,0)$) {\begin{varwidth}{1.5cm}Coordinated checkpointing\end{varwidth}}; 
      \node[draw,fill opacity=0.5,fill=gray] at ($(ucp) +(-0.5,-1.75)$) (lb-r) {\begin{varwidth}{1.5cm}Log-based rollback\end{varwidth}};
      
    \node[draw,fill=brown,fill opacity=0.5] at ($(ccp) +(-0.5,-0.75)$) (g-r) {\begin{varwidth}{1.5cm}Global rollback\end{varwidth}};
    \node[below of=g-r, node distance=2.5cm] (dfr){};
    \draw  (lb-r) edge node{\cite{Gamell2017}} (iccp);
    \draw  (lb-r) edge node{\cite{Guermouche2011}}(ucp);
    
    \draw  (g-r) edge node{\textbf{standard}} (ccp);
    \draw[dotted,thick]  (g-r) edge node[,fill=brown!50!gray,fill opacity=0.5]{partial \cite{Ropars2011}} (lb-r); 
    
    \draw[dotted,thick]  (dfr) edge[<->] node[draw,rotate=90,fill opacity=0.5,fill=white!50!yellow]{\begin{varwidth}{2.5cm}DFR\end{varwidth}} (g-r);
    \draw(dfr) edge node[label={[rotate=55] above:this work}]{} (iccp);

    }
\end{tikzpicture}

\caption{A graphical presentation of related work in rollback recovery strategies, considering message logging overhead, and recompute from rollback, as $x$ and $y$ axis. Data-flow-driven recovery, as proposed in this work, has no logging overhead, and its rollback depends on the data-flow dependencies of the underlying kernel.}
\label{fig:related-work}
\end{figure}

A comprehensive survey of rollback recovery strategies, divided into checkpoint-based and log-based, is given by Elnozahy et al. \cite{Elnozahy2002}; our work fits into checkpoint-based rollback, but requires application modifications, which is beyond the main focus of the survey.
Regardless of the chosen rollback, an advanced MPI implementation is indispensable when using and experimenting with various recovery options; we use ULFM \cite{Bland2013} to implement data-flow recovery mechanisms for MPI codes.

We provide an up-to-date and graphical outline of closely related global and non-global rollback techniques in Fig. \ref{fig:related-work};
our work is positioned within the related work as well.
The two dimensions we study are message logging, and the level of rollback (global or non-global). 

The most popular approach to resilience is coordinated checkpointing, combined with global rollback.
This mechanism requires no message logging, and guarantees that a rollback to a globally consistent checkpoint is done across any application codes.
In this approach, all processes (synchronously or asynchronously) write local checkpoints, which build a consistent global checkpoint.
In case of a process failure, all processes roll back to this checkpoint, and they lose the progress made since that checkpoint.
The advantage of this approach is that it is robust, it requires no execution logs, and the overhead in programming it is manageable (even if not trivial).
However, global rollback potentially wastes the progress across many running processes, especially in large-scale parallel runs.

A different solution to this problem is in the use of uncoordinated log-based protocols, which target the reduction of required rollback.
A runtime, such as a modified MPI library, can log messages exchanged between processes, and selectively take checkpoints at each process (uncoordinated checkpointing).
In the case of a failure, processes follow one of a number of message-logging protocols \cite{Alvisi1998} to roll back to the last globally consistent cut, with the support of available message logs.
In the general case, such rollback is not guaranteed to work; the so called ``domino effect'' may roll back all participating processes to the start of the application \cite{Elnozahy2002}.
Variations on log-based rollback mechanisms exist to resolve some of the existing issues.
The promise of these is the reduction of rollback and uncoordinated checkpointing.
The main challenges, however, are the logging overhead and the complexity of protocols.

One way to address this issue is to focus on a subclass of applications; a recent contribution \cite{Guermouche2011} implements a complex log-based protocol in an MPI runtime, which performs local recovery for a large set of MPI applications (send-deterministic applications).
Another way is to focus on an application, and implement application-specific log-based protocols as is the case for the S3D application \cite{Gamell2017}.
Compromises between the extremes of global rollback and log-based rollback, called hybrid or partially message-logging protocols, also exist \cite{Ropars2011}.

Within the existing work, the proposed data-flow driven rollback fits as shown in Fig. \ref{fig:related-work}: 
DFR, unlike all log-based approaches, does not log messages, being based on data flow analysis of the application kernel and programming the localised rollback at program level.
In this respect, it is similar to global rollback, which requires no message logging.
However, DFR can function as a localised rollback, as we have shown in this work, as long as the DFG of the kernel show local dependencies.


In our work, we are not interested in data flow dependencies within a process, obviously a central topic of interest in the compilers domain.
Instead, we study data flow dependencies between MPI processes.
Some research exists on statically analysing the data flow of MPI applications \cite{Strout2006}, based on the MPI communication calls of the codes.
However, data flow and its role in implementing reduced rollback strategies for distributed codes, has not been studied and advocated, to the best of our knowledge, in the MPI domain.
There is a certain duality between the studied data flow properties statically, and their manifestation at runtime as communication activities.
A study of communication activities for various HPC codes has been done by Kamil et al.~\cite{Kamil2005}.

In a study on energy consumption for fault tolerance, Meneses et al.~\cite{Meneses2014} also focus on rollback to improve energy efficiency.
The authors parallelise the global rollback efficiently in AMPI; in contrast, we design non-global rollback, thereby reducing its amount.

Aside from the MPI domain, task-based runtimes, which have an explicit model of task and data dependencies, have explored localised rollback. 
For example, localised rollback has been proposed for the Kaapi framework \cite{Besseron2008}.
More recently, a distributed version of Cilk designed an efficient scheme for localised recovery \cite{Kestor2017} in fork-join programs; as all Cilk programs, these rely on the master-slave paradigm, and focus on recovering work units (such as Fibonacci), but not global data, for distributed-memory runs.

Simulation has been used to explore fault tolerance in the past.
Some highly scalable MPI simulators have been extended for resilience studies~\cite{Levy2013,Naughton2014}; alternatively, special-purpose resilience simulators for specific studies~\cite{Ferreira2011,Gamell2017} have been developed.
With the exception of LogGOPSim  \cite{Levy2013}, which is actively maintained, these simulators are not available to the community to the best of our knowledge.
LogGOPSim is a trace-driven simulator, while SimGrid is execution-driven.
Trace-driven simulators directly, and thus more accurately, capture the underlying experimental platform, while execution-driven simulators such as SimGrid need to model the underlying platform, for example via a configuration file.
However, execution-driven simulation does offer advantages in the ease and flexibility of testing various scenarios, and the detachment from an application or MPI library implementation.

\section{Conclusion and Limitations}
\label{sec:conclusion}

In this manuscript, we introduced a version on non-global rollback, which differs from the related work in the MPI domain, since it is not based on message logging, but on a careful analysis of the data flow graph of an application kernel.
The analysis enables a localised rollback scheme for many applications.
We have demonstrated this on the example of the popular Jacobi method.
In order to quantify the end-to-end benefits of this approach, we combined the localised rollback with a CPU frequency scaling technique.
This resulted in reproducible energy savings of 10-12\% of the machine power consumption (10-15 J/s) for all idle processes, and for the duration of the rollback.
A simple model allowed us to estimate that the energy savings, as opposed to global rollback, scale as $n^2$, if $n$ is the MPI process count, when DFR is applied to stencils.
One factor is due to the increase of failures across the system with increase in nodes.
The second factor, which applies only for codes with localised dependencies (such as stencils), is due to the localised impact of each failure.
Significantly, the rollback via data-flow analysis requires no message logging, and has similar benefits to the state-of-the-art localised rollback methods employing logging.

One notable disadvantage of our work, compared to a generic log-based localised rollback \cite{Guermouche2011}, is that each data-flow recovery needs to be designed with an application kernel in mind; however, since data flow dependencies are more abstract than kernels, it can be argued that each design could cover a class of applications, as is the case for neighbourhood dependencies, and stencils.
Another challenge is that applications with global data-flow dependencies at each iteration step, such as illustrated in Fig. \ref{fig:intro}(b), are most likely to produce a global rollback scheme.
However, related work employing message logging for such kernels similarly shows prohibitively large runtime overheads.
Therefore, this problem may point to the general issue of applying any localised rollback for certain tightly coupled kernels, and does not disprove the use of the proposed data-flow rollback scheme.

We also faced various technical difficulties, both in the MPI implementation and the used simulator.
On the one hand, while ULFM is an MPI library with advanced fault tolerance features, we did experience bugs (e.g. even when scaling up failures using the original recovery scheme).
On the other hand, SimGrid also showed unexpected bugs during partial replacement of the execution DAG.
The scalability of SimGrid was impressive, handling millions of tasks in less than an hour, improving significantly on a much slower vanilla implementation based on Python; however, these settings are still very limited -- a million tasks are already contained in a DAG representing 1000 MPI processes for 1000 iterations.
Therefore, we are unable to run simulations of larger scale at this moment.
Another challenge we faced was the difficulty of performing fine-grained power measurements with good visual tools; in the end, we had to devise our own experimental setting and visualisation.
Still, when benchmarking overall runtime and performance, it is difficult to distinguish between the performance of the fault tolerance functionality, and the performance of the essential application code; a deeper analysis of power consumption without more advanced introspective tools would be difficult.

\section*{Acknowledgements}
This project has received funding from the European Union's Horizon 2020 research and innovation programme under grant agreement No 671603.
\bibliographystyle{acmart}
\bibliography{references}
\end{document}